\documentclass[a4paper,11pt]{article}
\usepackage{pos}
\usepackage{blindtext}
\usepackage{url}
\usepackage{mathtools}
\usepackage{bbm}
\usepackage{xcolor}
\usepackage{bbold}
\usepackage{caption}
\usepackage{subcaption}
\usepackage{wrapfig}
\usepackage{fancyhdr}
\usepackage{physics}
\usepackage{slashed}
\usepackage{appendix}
\usepackage{pifont}
\usepackage{float}
\usepackage{afterpage}
\usepackage{bm}
\usepackage{sidecap}
\title{Approach to meson-baryon femtoscopy using effective field theories}

\author*[a]{Álvaro Peña Almazán}
\author[b]{Juan M. Torres-Rincon}

\affiliation[a]{Facultat de Física, Universitat de Barcelona, Martí i Franquès 1, 08028 Barcelona, Spain}

\affiliation[b]{Departament de Física Quàntica i Astrofísica and Institut de Ciències del Cosmos (ICCUB), Facultat de
Física, Universitat de Barcelona, Martí i Franquès 1, 08028 Barcelona, Spain}

\emailAdd{apenaalm17@alumnes.ub.edu}
\emailAdd{torres@fqa.ub.edu}

\abstract{We calculate several femtoscopy correlation functions in the strangeness sectors $S=0$ and $S=-2$ for meson-baryon interactions. We combine the amplitudes of chiral perturbation theory at leading order with the TROY (T-matrix-based Routine for HadrOn femtoscopY) framework. We consider the correlation function for the $\pi^+$p and $\pi^-$p channels, which are currently under analysis by the ALICE collaboration at the LHC. Furthermore, we will show that analogous interactions can be used to reproduce the results for $K_S \Lambda$ correlation functions obtained by ALICE collaboration in PbPb collisions.}

\FullConference{%
  QNP2024 - The 10th International Conference on Quarks and Nuclear Physics, \\
   8-12 July, 2024 \\
   Barcelona, Spain 
}

\begin{document}

\maketitle

\section{Introduction}
In the field of relativistic heavy-ion collisions (RHICs), femtoscopy studies have been employed for various purposes throughout its history~\cite{Lisa:2005dd}. Recently, femtoscopy correlations have been used for investigating the strong interaction between hadrons and the possible dynamical states that can arise from them. The central quantity of femtoscopy analyses is the correlation function, which is defined as the ratio between the two-particle spectra and the product of two single particle distributions as functions of the relative momentum~\cite{Fabbietti_2021}.
As a theoretical approach to the correlation function we use the Koonin-Pratt formula~\cite{Lisa:2005dd,Fabbietti_2021},
\begin{equation} \label{Koonin-Pratt}
    C(\bm{q})=\int{}d^3r\sum_i \omega_i{} \ S_i(\bm{r})|\Psi_i(\bm{q},\bm{r})|^2 \ ,
\end{equation}
where $\bm{q}$ is the relative momentum of the two particles in the center of mass (CM) frame and $\omega_i$ are weights of the hadron pair $i$ that measures how abundantly it is produced in the collision. $S_i(\bm{r})$ is the source function, which represents 
the probability of a pair $i$ to be emitted at a relative distance $\bm{r}$. Finally, $\Psi_i(\bm{q},\bm{r})$ is the pair wave function that will relate the initial channel, $i$, to the asymptotic final one. The sum will extend over all the channels $i$ that couple to the asymptotic final one. 

For the source function---which we assume to be independent of $i$---we follow previous works~\cite{Fabbietti_2021,LambdaKAlice} and adopt a spherically symmetric Gaussian form depending only on the Gaussian radius $R$. In this work, we use different values for $R$ depending on the channel we study, and the type (pp or nucleus-nucleus) and centrality of the collision~\cite{LambdaKAlice}.

To obtain the wave function of the meson-baryon pair, $\Psi_i(\bm{q},\bm{r})$, we model the interaction using baryon chiral perturbation theory ($\chi$PT)~\cite{HYODO2012} and consider only contact terms (Weinberg-Tomozawa interaction) projected onto the $s$-wave~\cite{HYODO2012},
\begin{equation} \label{potencial fuerte $S$-wave}
    V_{ij}^{s-\textrm{wave}} =-\frac{C_{ij}}{4f_if_j}\sqrt{\frac{E_j+M_j}{2M_j}}\sqrt{\frac{E_i+M_i}{2M_i}} \left(2\sqrt{s}-E_i-E_j +\frac{p_i^2}{E_i+M_i}+ \frac{p_j^2}{E_j+M_j} \right) \ ,
\end{equation}
where $p_i$, $p_j$ are the CM relative momenta in the initial and final states, $C_{ij}$ are real isospin coefficients and $f_i$, $f_j$ the meson decay constant of the initial and final channels. The masses and energies refer to the baryons of each channel.

To overcome the low-energy limitations of $\chi$PT, we will introduce a unitarized version of the amplitudes, obtained from the solution of a Bethe-Salpeter (BS) equation. The nonperturbative equation leading to the $T$-matrix for the meson-baryon interaction reads~\cite{HYODO2012},
\begin{equation}\label{BS integral}
    T_{ij}(p_i,p_j;P)=V_{ij}(p_i,p_j;P)+i\sum_{l}\int\frac{d^4q}{(2\pi)^4}V_{il}(p_i,q;P)\mathcal{D}^\mathcal{B}_l(q;P)\mathcal{D}^\mathcal{P}_l(q;P)T_{lj}(q,p_j;P),
\end{equation}
where $P$ is the total four momentum, $V_{ij}$ is the interaction potential between channels $i$, $j$ and $\mathcal{D}^\mathcal{B}_l(P-q), \ \mathcal{D}^\mathcal{P}_l(q)$ are the baryon and the pseudoscalar meson propagators for the channel $l$.

\section{TROY framework}

The calculation of the $T$-matrix elements~\eqref{BS integral} will be performed using TROY ($T$-matrix-based Routine for hadrOn femtoscopY). Using the interaction potential of Eq.~\eqref{potencial fuerte $S$-wave}, TROY first computes the off-shell \( T \) matrix elements given in (\ref{BS integral}), using a Gaussian form factor, $V_{il}(p_i,q;P)\rightarrow V_{il}(p_i,q;P)\exp(-(p_i^2+q^2)/\Lambda^2)$, to regulate the momentum integral, where $\Lambda$ is a ultraviolet cut-off which is determined by fixing the pole positions of the generated resonances to their physical values, as we will indicate later.
Once the off-shell $T$ matrix is calculated, TROY computes the \( s \)-wave projection of the wave function, necessary for the correlation function~\cite{JuanDmesons},
\begin{equation}
\varphi_i(q;r)=j_0(qr)\delta_{if}+\int^{\infty}_{0}\frac{4\pi q^{\prime 2}dq^\prime}{(2\pi)^3}\frac{T_{if}(q^\prime,q;\sqrt{s})j_0(q^\prime r)}{2E_{\mathcal{B},i}2E_{\mathcal{P},i}(\sqrt{s}-E_{\mathcal{B},i}-E_{\mathcal{P},i}+i\eta)} \ . \label{eq:wavefunction}
\end{equation}
As some channels involve charged particles ($\pi^-p,\ \pi^+p$), the Coulomb interaction is included into the potential for a complete description of the wave function. See technical details in Ref.~\cite{JuanDmesons}.

Assuming that only the strong $l=0$ partial wave affects appreciably the wave function, the Koonin-Pratt formula, Eq.~\eqref{Koonin-Pratt} can be written as
\begin{equation}
C(q)=\int{}d^3rS(r)|\Phi^C_f(\bm{q};\bm{r})|^2+\int{}4\pi{}r^2drS(r)\left[\sum_i\omega_i|\varphi_i(q;r)|^2-|\Phi^C_{0f}(qr)|^2\right] \ ,
\end{equation}
where $\varphi_i(q;r)$ is the total wave function that includes the $s-$wave strong and Coulomb interactions, $\Phi^C_f(\bm{q};\bm{r})$ is the total Coulomb wave function (with all partial waves included) and $\Phi^C_{0f}(qr)$ is its $s-$wave projection.

\section{Results: meson-baryon correlation functions}

Experimental data for $\pi^-p$ and $\pi^+p$ channels are currently under analysis by the ALICE collaboration~\cite{Lesch}. We have implemented the meson-baryon potentials for the $S=0$ sector and used a cutoff of $\Lambda=1125 \ \mathrm{MeV}$, which reproduces previous results using the on-shell $T$-matrix equation to fix the $N(1535)$ resonance~\cite{Sun_2019}. Details on the methodology will be given elsewhere~\cite{Peña}.

\begin{figure}[H]
    \centering
    \includegraphics[width=0.46\linewidth]{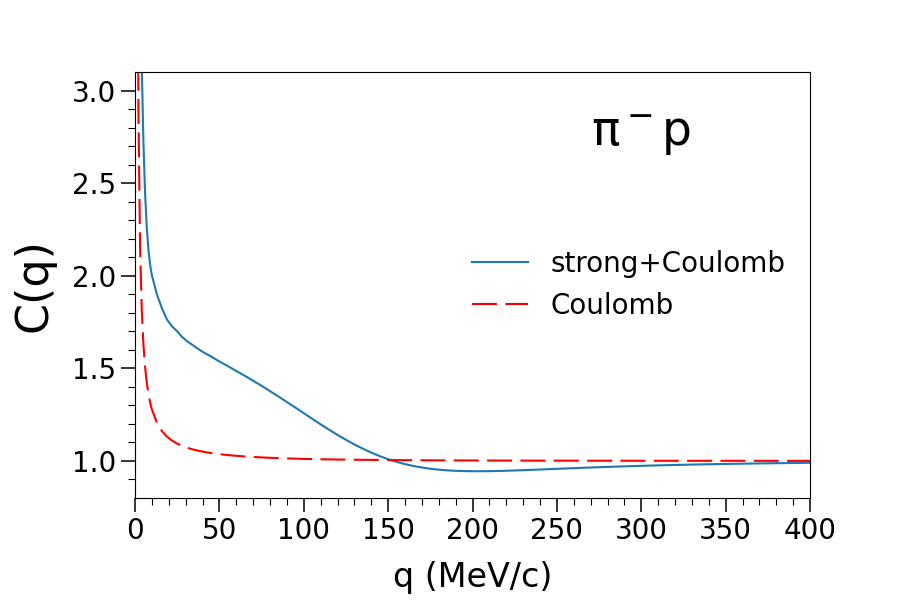}
    \includegraphics[width=0.46\linewidth]{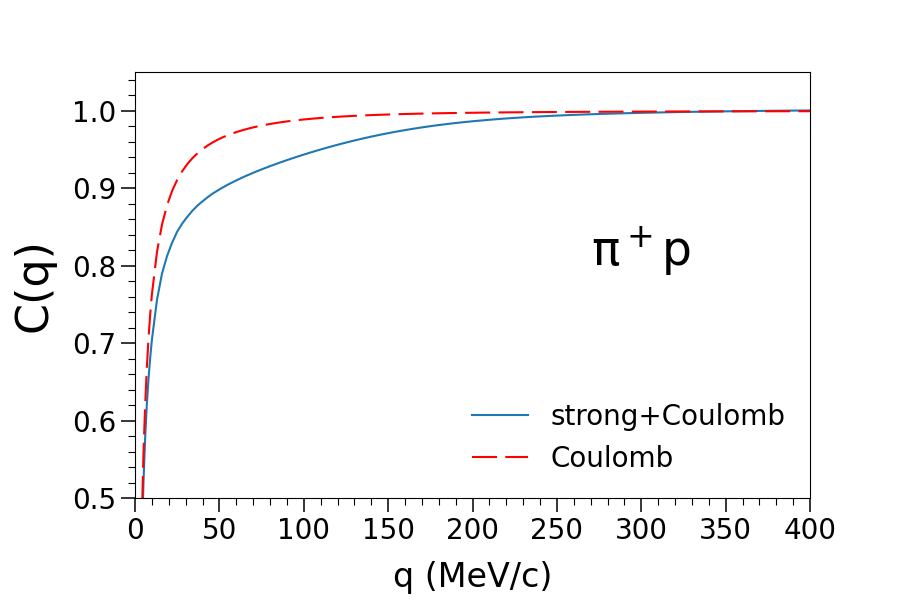}
    \label{PimP radius}
    \caption{Correlation functions for the $\pi^- p$ $(S,Q)=(0,0)$ (left) and $\pi^+p$ $(S,Q)=(0,2)$ (right) as functions of the relative momentum. We use a Gaussian radius of $R=1$ fm, a typical value for pp collisions.}
    \label{Corr_pip}
\end{figure}

The results for the $\pi^-p$ channel are shown in the left panel of Fig.~\ref{Corr_pip}. The effect of coupled channels is added to the final correlation function, where we used $w_i=1$ for simplicity. The correlation function remains mostly above one, reflecting the attractive influence of the Coulomb force. The modification of the strong interaction is large and positive, given the attractive character in this channel. In contrast, the $\pi^+p$ case, shown in the right panel of Fig.~\ref{Corr_pip}, exhibits a repulsive Coulomb contribution that generates a correlation function below one. Additionally, in this channel, the strong interaction is also repulsive, as shown by the fact that the curve with both interactions is below the curve with only the Coulomb contribution.

While in the $T$-matrix equation we generate the $N(1535)$ in both channels, this resonance has no visible effects on the correlation functions, as the $\pi N$ coupling to this resonance is rather small~\cite{Sun_2019}. It is important to note that we do not account for the $p$-wave $\Delta$ resonance, making our results to lack of an important contribution which will be accommodated in the future. So far, our results are only reliable after experimental filtering of the $\pi N$ pairs that come from the $\Delta$ resonance.

Finally, we focus on the correlation function of the $K_S\Lambda$ system, in which we require to combine two different correlation functions, since $\ket{K_S}=\frac{1}{\sqrt{2}}\left(\ket{K^0}+\ket{\bar{K}^0}\right) \ $. To solve the $T$-matrix equation for the $\bar{K}^0\Lambda$ channel, belonging to the $S=-2$ sector, we have used $\Lambda=770 \ \mathrm{MeV}$ as a natural value, where the scale is taken as the $\rho$ meson mass. Details will be given in Ref.~\cite{Peña}.

\begin{figure}[H]
    \centering
    \includegraphics[width=0.7\linewidth]{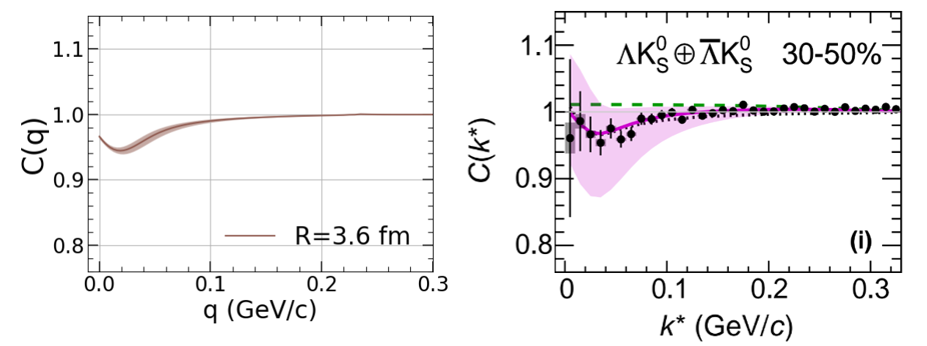}
    \includegraphics[width=0.45\linewidth]{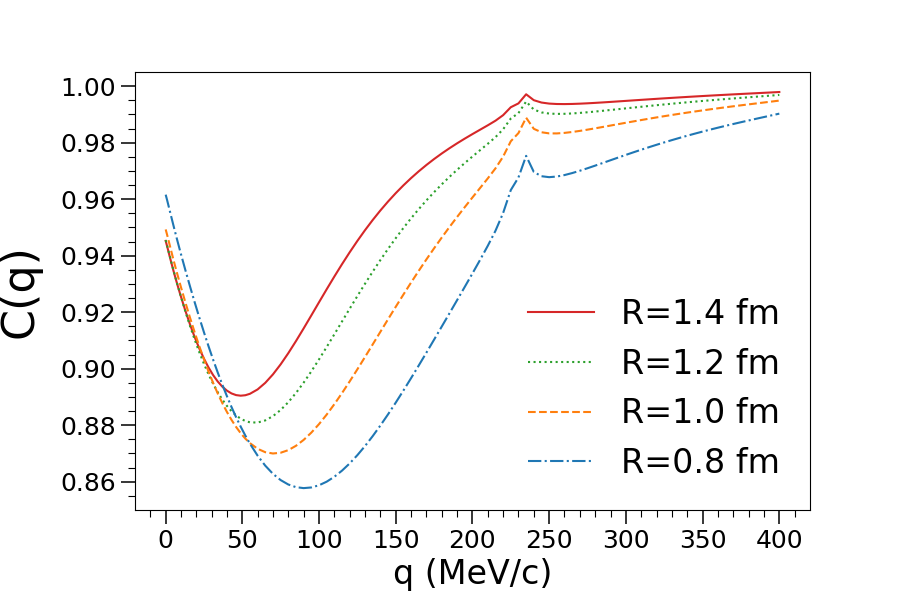}
    \caption{Correlation function for the $K_S\Lambda$ system. Top, left: TROY calculation using baryon $\chi$PT with $R=3.61$ fm. Top, right: Experimental result from ALICE using PbPb mid-peripheral collisions at $\sqrt{s_{NN}}=2.76$ TeV~\cite{LambdaKAlice}. Bottom: TROY prediction for pp collisions using small Gaussian radii.}
    \label{K_S Comparison}
\end{figure}

The result for the $K_S \Lambda$ correlation function is shown in Fig.~\ref{K_S Comparison} (top, left) compared to ALICE results in mid-peripheral PbPb collisions (top, right). To account to the larger size of that system we applied $R=3.61(44)(30)$ fm~\cite{LambdaKAlice} in the theoretical prediction. Despite the rather flat structure of the correlation function we observe an excellent match between our result and the experimental data. Our error band is coming from the uncertainty of the Gaussian radius. The minimum that appears in our calculation is due to the combined effect of two resonances: the $N(1535)$ in the $K^0\Lambda$ channel ($S=0$) and the $\Xi(1620)$ which appears in the $\bar{K}^0\Lambda$ channel ($S=-2$).

In the bottom panel of Fig.~\ref{K_S Comparison} we present the $K_S \Lambda$ correlation function for pp collisions calculated by reducing the Gaussian radius to the range $R \in (0.8,1.4)$ fm. We observe the minimum produced by the $N(1535), \Xi(1620)$ resonances as well as a cusp due to the opening of several $K\Sigma$ channels.

\section{Conclusions and Outlook}

We have presented our first results on meson-baryon femtoscopy correlation functions in the $S=0$ and $S=-2$ sectors using the TROY formalism. From effective potentials calculated at leading-order in baryon $\chi$PT we have solved the $T$-matrix equation and reconstruct the meson-baryon wave functions. Finally, these have been implemented into the Koonin-Pratt formula to obtain the correlation functions of $p\pi^+, p\pi^-$ and $K_S \Lambda$. We have covered pp and PbPb collisions measured by the ALICE collaboration, by varying the range of the source function. In the future we plan to provide other meson-baryon channels, adjust the weighting factors $w_i$ (so far taken to one), and incorporate the $p$-wave narrow states, like the $\Delta$ and $\Xi(1680)$ resonances, which are expected to have a prominent role in the final correlation functions.

\acknowledgments

This work has been supported by the project number CEX2019-000918-M (Unidad de Excelencia ``Mar\'ia de Maeztu''), PID2020-118758GB-I00 and PID2023-147112NB-C21, financed by the Spanish MCIN/ AEI/10.13039/501100011033/.
AP thanks the Institute of Cosmos Sciences (ICCUB) for financial support during the master's degree.

\end{document}